# Algebra of hypersymmetry
(extended version)
## applied to state transformations in strongly relativistic interactions
illustrated on an extended form of the Dirac equation

György Darvas

Symmetrion, Budapest, Hungary,
ORCID 0000-0002-1652-4138
E-mail: darvasg@iif.hu

**Abstract**: There are several 3 + 1 parameter quantities in physics (like vector + scalar potentials, four-currents, space-time, four-momentum, …). In most cases (but space-time), the three- and the one-parameter characterised elements of these quantities differ in the field-sources (e.g., inertial and gravitational masses, Lorentz- and Coulomb-type electric charges, …) associated with them. The members of the field-source pairs appear in the vector- and the scalar potentials, respectively. Sections 1 and 2 of this paper present an algebra what demonstrates that the members of the field-source siblings are subjects of an invariance group that can transform them into each other. (This includes, e.g., the conservation of the isotopic field-charge spin (IFCS), proven in previous publications by the author.) The paper identifies the algebra of that transformation and characterises the group of the invariance; it discusses the properties of this group, shows how they can be classified in the known nomenclature, and why is this pseudo-unitary group isomorphic with the SU(2) group. This algebra is denoted by tau ($\tau$). The invariance group generated by the tau algebra is called hypersymmetry (HySy). The group of hypersymmetry had not been described. The defined symmetry group is able to make correspondence between scalars and vector components that appear often coupled in the characterisation of physical states. In accordance with conclusions in previous papers, the second part (Sections 3 and 4) shows that the equations describing the individual fundamental physical interacions are invariant under the combined application of the Lorentztransformation and the here explored invariance group at high energy approximation (while they are left intact at lower energies). As illustration, the paper presents a simple form for an extended Dirac equation and a set of matrices to describe the combined transformation in QED. The paper includes a short reference illustration (in Section 2.2) to another applicability of this algebra in the mathematical description of regularities for genetic matrices.

**Keywords**: transformation matrix, symmetry group, hypersymmetry, invariance, SU(2), 3+1 parameter quantities, isotopic field-charge, isotopic field-charge spin, conservation, invariance

PACS numbers: 02.20.Rt, 11.30.Ly, 11.10.-z, 12.20.Ds, 11.15.Kc

## Introduction

*Retrospective remarks*

In order to highlight the physical background that demanded to elaborate hypersymmetry (HySy), one needs to make acquainted with a few notions first. The physical notion of the *isotopic field-charge*s (*IFC*)[1] was introduced in [11]. There was shown that this physical quantity has a

---
[1] Definition: Similar to the distinction between gravitational and inertial masses that appear as sources of the gravitational and kinetic (scalar and vector) parts in the Hamiltonian of a mechanical system, respectively, one may make distinction between field-charges of other interaction fields as well. If one considers the *gravitational and inertial masses as isotopic states of mass* (as field-charge of the gravitational interaction field), in an analogical way, one can define the field-charges appearing in the roles of sources in the scalar (potential) and vector (kinetic) parts of

property, called *isotopic field-charge spin* (IFCS, denoted by $\Delta$) that (a) can be rotated in an abstract gauge field; (b) can take two values during this transformation; and (c) may transform the two isotopic states of any field-charge into each other. $\Delta$ proved to be a conserved quantity. The proof [11] was demonstrated *analitically*, deriving conserved currents from general field equations (not specified to any concrete interaction field) in the presence of a kinetic gauge field.[2]

There was left open to derive and discuss *algebraically* the properties of a group that charcterises the conservation of $\Delta$. The group that transforms the IFC-s into each other has still not been described in a complete form for physics.

The isotopic field-charge (IFC) theory takes into account *sources* of respective *scalar* and *vector fields* as *different* physical properties. These two kinds of sources are considered as *isotopic states* of each other that are, at least at rest, quantitatively equivalent, but qualitatively different. Putting them in the formula of field-charge currents, they would destroy those invariance under usual transformations, unless *an additional invariance*, that can transform the two isotopic states into each other, restores the apparently "lost" order[11] [3]. The existence of such invariance has been *analytically* proven. It was the conservation of a property of IFC-s: the so called *isotopic field-charge spin* ($\Delta$).

*What will be derived in this paper?*

Now, we explore the *algebra of this additional invariance* and *the respective group* that transforms three-component columns into scalars, and vice versa. This group is able, among others, to transform the isotopic states of field-charges of any fundamental physical interaction field into each other. As a consequence, equations describing fundamental physical interacions, at least at high energies, should be invariant under the combined application of the Lorentz-transformation and this additional invariance group. Similar (slightly different) ideas and approaches – that are refered to and discussed in this paper – have been partly confirmed and partly based by a series of cited papers that are (direct and indirect) signs of the actuality of the discussed topic, among others, that led to the need for high energy extensions of the Dirac equation. (Cf., [17], [25][4], [24], [19], [26] and recently [36], [37]).

Since those isotopes of the field-charges that are sources of vector-fields appear in the field-charge currents as coefficients of three-velocities, the matrices in the sought algebra of their transformation into isotopes of the field-charges of the respective scalar field, and *vice versa*, should take the form of [4x4]. There must appear 3-row minors of coinciding elements in these matrices that act, among others, on those isotopic field-charges that appear as multipliers of the three velocity vector components in the charge currents.

*The paper demonstrates that these matrices form a group. It discusses the properties of this group, shows how they can be classified in the known nomenclature, why this pseudo-unitary group is isomorphic with the SU(2) group, and gives examples for its applications.*

---

the Hamiltonian of any other interaction field also as isotopic states of each other. These isotopic state field-sources are called *isotopic field-charges* (*IFC*) of the given interaicition field.

[2] A recent classification of Noether-charges appeared in [46]; cf., also [10].

[3] Recently, a very similar paper [47] was published in *Eur. Phys. J. C.* (without reference to [10]) that discusses a Lorentz violating extension of a QED, derived also from electron-electron scattering. Their conclusion leads also to a gauge invariant model.

[6] discussed another Lorentz invariance breaking interaction model that violates Lorentz (and CPT) symmetry at very high energies.

[4] Goenner [25] discussed a more general approach to extended Lie algebras that can be applied in gauge fields.

The pseudo-unitary properties of the discussed group are described in [38], [35], physical applications of such pseudo-unitary groups were discussed in detail repeatedly by [29], [30], [31], [32], [33] and [1], [2], [3][5], as well as Hermiticity was discussed especially by [7], and there are all applied below in Sections 1.2-1.4.

The cited theory, which the presented algebra is applied to, takes into consideration the transformation of *the velocity* and of *the respective field-charges* separately. Therefore, in contrast to the phase space of classical quantum field theories, it applies description in the configuration space. Thus, velocities are transformed in *space-time*, while the field-charges (like mass) are subjects of transformation in a respective kinetic (velocity dependent) *field*. The latter (as *field*) has been introduced in the IFCS theory. This separation is in compliance with a property of the Lorentz-group ($SO^+(3,1)$) that splits in two subgroups, namely to [$R(\varphi)$] that characterizes rotations in the space-time, and [$\Lambda(\dot{x}_\nu)$] that characterizes velocity boosts into a given direction in the configuration space (cf., Section 5).

The current paper presents an algebra of transformation matrices of the invariance group that may describe the isotopic field-charge spin (IFCS or $\Delta$) conservation, in a more exact and extended way than the earlier published analytical proof. In accordance with the conclusions in [11], the equations describing the individual fundamental physical interacions are invariant under the combined application of the Lorentz-transformation and this invariance group (Sections 1 and 2) at high energies.

As an application of this (3 + 1) algebra[6], the next part of the paper demonstrates its applicability on the example of an extended form of the Dirac equation (Sections 3 and 4). Previous papers [12], [14] demonstrated that this extension causes deviations from the results derived in the Standard Model at very high energies (at velocities close to that of the light); while the explored additional symmetry is broken at lower (cca. <0.9c) energies. Thus, the SM holds asymptotically at 'low' energies. The Dirac equation was 'weekly-relativistic' (according to Dirac an approximation, c.f., his introductory remarks in [20]) in the sense that it was invariant under the Lorentz-transformation, at least at not too high velocities, however it could not handle 'strongly-relativistic' interactions in "the theory of the electron" at extremely high velocities. Further attempts to cure this incompleteness were made by Dirac [22], [23] and others in the course of the past nine decades. Approaches by the use of Finsler-metrics was attempted by several authors, (among others, a similar was proposed by Dirac, 1962 without mentioning Finsler); we refer here only to three papers [34], [4], [5, Sec. 4]. The author of this paper has made also attempts in this direction [12], [14]. The illustrative Part (2) of this paper tries to add a "brick" to those efforts.[7]

---

[5] Ahmed and Jain [3] included also singular matrices, and applications to Klein-Gordon-type equations.

[6] [18] presents a symplectic Hamiltonian reduction for 3+1-dimensional free and Abelian, massive and massless, gauged Rarita–Schwinger theories affecting fermionic fields.

[7] Some ask, what is the relation of this approach to the Standard Model? Well, a few parallel working groups discussed the possible issues of the necessary extension towards a New Physics beyond the SM in the years 2005-2008. (A few authors discussed the need for a "new physics" already in the late nineties.) I refer only to the report by the CERN *Working Group 3* in this series [9] that demonstrated (among others) that the expected New Physics would not work on the basis of electric charge-charge interactions based only on the "classical" QED. The expected New Physics is not a product of dreaming fantasy. It is partly based on experimental data partly it extends the SM at extreme conditions, while keeping its validity among those conditions for which that was elaborated. The cited report is based on the SUSY, while my approach (HySy) belongs to the series of nearly a dozen alternatives of the SUSY model. A group in the CERN is looking for so called "exotic" particles among the LHC collision products' data, which could show signs of non-supersymmetric new particles beyond the SM. Discussion over them is open in series of conferences and journals. Further experiments will test, which of them will be proven the best approximation of the reality at GUT conditions. This not-replacing-rather-extending-the-SM character of the aimed New Physics is important in the judgment of the "rank 2" of the $\tau$-matrices.

PART 1

**1 Matrix algebra for 3+1 parametric transformations**

We investigate matrices (denoted by $\tau$) that can function as operators. These operators effect state functions. One can define eigenvalue equations in which the effect of the $\tau$ operators is multiplication by numbers (called eigenvalues of the $\tau$ operator). The state functions in these eigenvalue equations take the form of eigenfunctions of the $\tau$ operators. Since we want to represent the linear operator $\tau$ in form of square matrices, we are looking for the eigenfunctions in the form of eigenvectors. We would like to find such $\tau$ operators that are able to transform vierbeins characterised by 3+1 parameters into each other [48][8]. The operators $\tau$ should fulfil eigenvalue equations $\tau\varphi = k\varphi$, where $\varphi$ are eigenfunctions of the operator $\tau$, and $k$ are numbers, eigenvalues of this equation. As *Theorem 4* below will show, the tau algebra will be unitary, we fix the egienvalues of $\tau$ by $\pm\frac{1}{2}$. We denote the $\tau$ operators' eigenfunctions belonging to the eigenvalues of $\tau$ by $\varphi_+^{(\tau)}$ and $\varphi_-^{(\tau)}$, respectively. For simplicity, we denote the vierbein forms of the latter two eigenvectors by $\chi$ and $\vartheta$, respectively (cf., e.g., Eq. (13)). Having known the eigenvalues and the eigenfunctions, we are seeking operators $\tau$ that fulfil the above eigenequation.

**1.1** *The* tau *algebra*

Let's choose two eigenvectors $\chi$ and $\vartheta$ [with (+ + + -) signature] in the following representation:

$$\chi = \begin{bmatrix} 1 \\ 1 \\ 1 \\ 0 \end{bmatrix}, \quad \vartheta = \begin{bmatrix} 0 \\ 0 \\ 0 \\ i \end{bmatrix} \tag{1}$$

Note, we had certain limited freedom to choose the values in the eigenvectors. This choice makes handling the $\tau$ transformation-matrices convenient. Recall, that in the following case, like in many other cases, the representation of the transformation group coincides with a representation of the respective Lie algebra.

We aim at finding matrices, which transform linearly these two eigenvectors that satisfy the eigenvalue equations for the operator $\tau$ as described below. Similar to the $\sigma$ algebra in Dirac's QED [20], we are looking for matrices $\tau$ with the following properties:

$$\tau_3 \chi = \chi \qquad \tau_3 \vartheta = -\vartheta$$
$$\tau_2 \chi = \vartheta \qquad \tau_2 \vartheta = \chi$$
$$\text{requiring } \tau_i \tau_j = i\tau_k \text{ } (i, j, k \text{ cyclic indices}):$$
$$\tau_1 \chi = -i\vartheta \qquad \tau_1 \vartheta = i\chi$$

A set of the following matrices meets the required properties:

$$\tau_1 = \begin{bmatrix} 0 & 0 & 0 & 1 \\ 0 & 0 & 0 & 1 \\ 0 & 0 & 0 & 1 \\ 1 & 0 & 0 & 0 \end{bmatrix}, \quad \tau_2 = \begin{bmatrix} 0 & 0 & 0 & -i \\ 0 & 0 & 0 & -i \\ 0 & 0 & 0 & -i \\ i & 0 & 0 & 0 \end{bmatrix}, \quad \tau_3 = \begin{bmatrix} 1 & 0 & 0 & 0 \\ 1 & 0 & 0 & 0 \\ 1 & 0 & 0 & 0 \\ 0 & 0 & 0 & -1 \end{bmatrix} \tag{2}$$

---

[8] In contrast to the Dirac bispinor consisting of 2+2 components (two-spinors), our spinors have a peculiarity that must be reflected in their construction. Since they are expected to make (in classical terms unusual) correspondence between *vector components* and *scalars* (c.f., also Sec. 1.4), they must be constructed in a 3+1 form. Four-columns of these bispinors can be subjects of linear transformation by [4x4] matrices whose minors should reflect that 3+1 structure. This predicts the structure and explains the character of the sought $\tau$-matrices.

These $\tau$ matrices satisfy the following algebra:

$$\tau_1^2 = \tau_2^2 = \tau_3^2 = \begin{bmatrix} 1 & 0 & 0 & 0 \\ 1 & 0 & 0 & 0 \\ 1 & 0 & 0 & 0 \\ 0 & 0 & 0 & 1 \end{bmatrix} = \mathbf{E}, \quad \{\tau_i, \tau_j\} = 0, \quad [\tau_i, \tau_j] = 2i\tau_k \tag{3}$$

Introducing the notation: $\begin{bmatrix} 1 & 0 & 0 \\ 1 & 0 & 0 \\ 1 & 0 & 0 \end{bmatrix} = \mathbf{I}_L, \quad \begin{bmatrix} 0 & 0 & 1 \\ 0 & 0 & 1 \\ 0 & 0 & 1 \end{bmatrix} = \mathbf{I}_R$ the above matrices can be written in the form:

$$\tau_1 = \begin{bmatrix} 0 & \mathbf{I}_R \\ 1 & 0 \end{bmatrix}, \quad \tau_2 = \begin{bmatrix} 0 & -i\mathbf{I}_R \\ i & 0 \end{bmatrix}, \quad \tau_3 = \begin{bmatrix} \mathbf{I}_L & 0 \\ 0 & -1 \end{bmatrix} \quad \text{where} \quad \tau_1^2 = \tau_2^2 = \tau_3^2 = \mathbf{E} = \begin{bmatrix} \mathbf{I}_L & 0 \\ 0 & 1 \end{bmatrix}. \tag{4}$$

We denote the (diagonal) identity matrix of the [4 x 4] matrices by [**1**]. The *unit matrix* of the $\tau$ matrix-algebra is denoted by **E.**

**1.2** *Group properties of the τ-matrices*

*Theorem 1*: The $\tau$ matrices generate a group.

*Proof*: The $\tau$ matrices satisfy the four group axioms. The products of the $\tau_i$, plus the unit matrix **E**, together with their conjugate and negated matrices form a closed group; they are associative; there exists an identity element of the *group*, which is the unit matrix **E** defined by (3); and since the squares of all $\tau_i$ produce the unit matrix, all $\tau_i$ are the inverse group elements of themselves, that means $\tau_i^{-1} = \tau_i$. Note that the identity element of the group in group-theoretical sense is not the identity matrix [**1**], it is the unit matrix **E** defined in (4).

□

This is the group of hypersymmetry (HySy). Representation of the transformation-group is determined by the algebra of the $\tau$-matrices.
(As an example, they transform the two states of the isotopic field-charges among each other by a rotation in an abstract field described in [11].)

The expressions (4) of the $\tau$ matrices show very similar form to the [2 x 2] Pauli matrices:

$$\sigma_1 = \begin{bmatrix} 0 & 1 \\ 1 & 0 \end{bmatrix}, \quad \sigma_2 = \begin{bmatrix} 0 & -i \\ i & 0 \end{bmatrix}, \quad \sigma_3 = \begin{bmatrix} 1 & 0 \\ 0 & -1 \end{bmatrix}. \tag{5}$$

Later, we will use the partial formal analogy in Section 2. However, there are differences as well: in contrast to the diagonal $\sigma$ matrices, our $\tau$-matrices manifest an apparent chiral character.

Observe, that all the above matrices of the $\tau$-algebra are singular. In matrix algebraic terms, in general, singular matrices have neither inverse, nor adjoint matrices! This limits their commutation in scalar multiplication with other matrices. Nevertheless, in group theoretical sense, they have inverse and unit elements, considering the expressions in (4). Inverse elements of the singular matrices constituting the group are interpreted in Penrose-Moore sense and are called pseudo-inverses.[9]

We will demonstrate this on the example of the [2 x 2] $\tau$ matrices introduced in (4). These matrices differ from the [2 x 2] Pauli matrices that the numbers *1* are replaced by the minor matrices $\mathbf{I}_R$ and $\mathbf{I}_L$ in the upper row. *In matrix algebraic terms* we should take into account the singularity of these minor matrices (and extend it to [4 x 4] matrices). *In group theoretical terms* there is not required that the constituents in the matrices, forming elements of a group, should appear only as numbers. Therefore, when investigating the group properties, we will not consider the special properties of $I_R$ and $I_L$.

---

[9] See more on pseudo-inverses also in [28].

*Theorem 2:* $\tau$ are self-adjoint matrices.

*Definitions*: Adjoint matrices are defined in two different ways. *In strict mathematical sense*, adjoint matrices to $A_{ij}$ (= Adj$A_{ij}$) are transposed matrices with signature $(-1)^{i+j}$ composed of the determinants of the minor matrices $A_{ji}$ belonging to the elements $a_{ij}$ of $A_{ij}$. This interpretation is used for the definition of the inverse matrix of $A_{ij}$. *In physics*, Hermitic matrices are interpreted as adjoints to $A_{ij}$ in the sense of transposed matrices composed of the complex conjugate elements $\overline{a_{ji}}$ of $A_{ij}$ (and are denoted by $A^+$). This interpretation is used in course of the transformation of scalar vector products, and for defining *self-adjoint* physical operators. The two interpretations lead to different adjoint matrices.

*Proof*: In the first sense, there holds for 2 x 2 matrices: Adj$M_{2x2}$ = $-M$ + Tr($M$)·[**1**], where Adj$M$ denotes the mathematical adjoint of $M$; Tr($M$) denotes the trace of the matrix; and [**1**] denotes the identity matrix.

Tr($\tau_1$) and Tr($\tau_2$) = 0, Tr($\tau_3$) = Tr($I_L$) – 1 = 0, that means
$\quad$ Adj$\tau_i = -\tau_i$ $\quad$ ($i$ = 1, 2, 3). $\hspace{4cm}$ (6)

This coincides with the properties of Pauli's $\sigma$ matrices. (Note, here we didn't consider the singularity of $I_L$ and $I_R$, since the applied definition for adjunction holds for all 2x2 matrices independent of the singularity of minor matrices appearing in their elements.)

In the Hermitic adjoint sense $\sigma_i^+ = \sigma_i$ and $\tau_i^+ = \tau_i$, that means, both are self-adjoint in physical terms.
$\hspace{15cm}$ ☐

*Theorem 3:* The determinants of the $\tau$ matrices coincide with those of Pauli's $\sigma$ matrices (-1).

*Proof*: Instead of reading the determinants from the form of the matrices in (4) (which would be – $I_R$, $-I_R$, $-I_L$, respectively, that are not numbers, what would be expected from determinants), let us apply the following definition: $M$·Adj$M$ = det$M$·$E$, where $E$ is the unit matrix of the $\tau$-algebra.

Considering also (6):
$\quad \tau_i$·Adj$\tau_i = -\tau_i^2 = -E$ $\quad$ (here $i$ is not a summing index), $\hspace{2cm}$ (7)
and comparing this with the definition, one can read from here that
$\quad$ det $\tau_i = -1$ ($i$ = 1, 2, 3), $\hspace{8cm}$ (8)
this property coincides again with that of the Pauli matrices.
$\hspace{15cm}$ ☐

*Theorem 4:* The $\tau$ matrices are unitary.

*Proof*: Similar to the Pauli matrices ($\sigma_i^+\sigma_i = \sigma_i\sigma_i^+ = $ [**1**]) the [2x2] tau matrices fulfil the (Hermitic sense) unitarity condition: $\tau_i^+\tau_i = \tau_i\tau_i^+ = E$, where $E$, in terms defined in Eq. (4), represents the unit matrix of the $\tau$ algebra. Since, similar to $\tau_i$, the [4x4] form of $E$ is singular, the group composed by the tau matrices will be called pseudo-unitary.
$\hspace{15cm}$ ☐

*Theorem 5:* The $\tau$ matrices are the inverses of themselves.

*Proof*: In order to check the above read (cf., proof of Theorem 1) (pseudo-)inverse of the $\tau_i$ matrices:

$$\tau_i^{-1} = \frac{\mathrm{Adj}\,\tau_i}{\det \tau_i} = \frac{-\tau_i}{-1} = \tau_i. \tag{9}$$

□

**1.3** *Representation of the group composed by the τ-matrices*

Similar to the Pauli matrices, let us compose a representation of the τ-algebra in the following way:

$$\mathbf{K}_+ = (\mathbf{K}_1 + i\mathbf{K}_2); \quad \mathbf{K}_- = (\mathbf{K}_1 - i\mathbf{K}_2); \quad \mathbf{K}_3 = \frac{1}{2}\tau_3 \tag{10}$$

where $\mathbf{K}_i = \frac{1}{2}\tau_i$

so

$$\mathbf{K}_+ = \begin{bmatrix} 0 & 0 & 0 & 1 \\ 0 & 0 & 0 & 1 \\ 0 & 0 & 0 & 1 \\ 0 & 0 & 0 & 0 \end{bmatrix} = \begin{bmatrix} 0 & \mathbf{I}_R \\ 0 & 0 \end{bmatrix}; \quad \mathbf{K}_- = \begin{bmatrix} 0 & 0 & 0 & 0 \\ 0 & 0 & 0 & 0 \\ 0 & 0 & 0 & 0 \\ 1 & 0 & 0 & 0 \end{bmatrix} = \begin{bmatrix} 0 & 0 \\ 1 & 0 \end{bmatrix};$$

$$\mathbf{K}_3 = \begin{bmatrix} 1/2 & 0 & 0 & 0 \\ 1/2 & 0 & 0 & 0 \\ 1/2 & 0 & 0 & 0 \\ 0 & 0 & 0 & -1/2 \end{bmatrix} = \begin{bmatrix} \mathbf{I}_L/2 & 0 \\ 0 & -1/2 \end{bmatrix} \tag{11}$$

where

$$[\mathbf{K}_3, \mathbf{K}_+] = \mathbf{K}_+; \quad [\mathbf{K}_3, \mathbf{K}_-] = -\mathbf{K}_-; \quad [\mathbf{K}_+, \mathbf{K}_-] = 2\mathbf{K}_3;$$

$$\mathbf{K}_+\mathbf{K}_- = \begin{bmatrix} 1 & 0 & 0 & 0 \\ 1 & 0 & 0 & 0 \\ 1 & 0 & 0 & 0 \\ 0 & 0 & 0 & 0 \end{bmatrix} = \begin{bmatrix} \mathbf{I}_L & 0 \\ 0 & 0 \end{bmatrix}; \quad \mathbf{K}_-\mathbf{K}_+ = \begin{bmatrix} 0 & 0 & 0 & 0 \\ 0 & 0 & 0 & 0 \\ 0 & 0 & 0 & 0 \\ 0 & 0 & 0 & 1 \end{bmatrix} = \begin{bmatrix} 0 & 0 \\ 0 & 1 \end{bmatrix} \tag{12}$$

$$\mathbf{K}_+ + \mathbf{K}_- = \tau_1; \quad \mathbf{K}_+ - \mathbf{K}_- = i\tau_2; \quad \mathbf{K}_+\mathbf{K}_- - \mathbf{K}_-\mathbf{K}_+ = \tau_3; \quad \mathbf{K}_+\mathbf{K}_- + \mathbf{K}_-\mathbf{K}_+ = \mathbf{E}$$

*Theorem 6*: $\mathbf{K}^2$ is Casimir invariant.

*Proof*: Composing the sum $\mathbf{K}^2 = \mathbf{K}_1^2 + \mathbf{K}_2^2 + \mathbf{K}_3^2 = \frac{1}{2}(\mathbf{K}_+\mathbf{K}_- + \mathbf{K}_-\mathbf{K}_+) + \mathbf{K}_3^2 = \frac{1}{4}(\tau_1^2 + \tau_2^2 + \tau_3^2) = \frac{3}{4}\mathbf{E}$

we get that $\mathbf{K}^2$ is proportional to the identity map (in our case the identity is represented by the unit matrix $\mathbf{E}$ of the group), it is a Casimir invariant that commutes with all the three generators of the representation: $[\mathbf{K}^2, \mathbf{K}_i] = 0$ ($i = 1, 2, 3$).

□

*Theorem 7*: The eigenvalues belonging to the $\mathbf{K}$ operators are half-integer.

*Proof*: We denote the weight of this representation belonging to the eigenvector $\chi$ by $k^{(\chi)}$ and that belonging to the eigenvector $\vartheta$ by $k^{(\vartheta)}$.

$$\mathbf{K}_3 \chi = \frac{1}{2}\tau_3 \chi = \frac{1}{2}\mathbf{I}_L \chi = k^{(\chi)} \chi; \quad k^{(\chi)} = \frac{\mathbf{I}_L}{2}$$

$$\mathbf{K}_3 \vartheta = \frac{1}{2}\tau_3 \vartheta = -\frac{1}{2}\vartheta = k^{(\vartheta)} \vartheta; \quad k^{(\vartheta)} = -\frac{1}{2} \tag{13}$$

□

The effects of the $\mathbf{K}_\pm$ operators on the eigenvectors are:
$$\mathbf{K}_+\chi = 0, \quad \mathbf{K}_-\chi = -i\vartheta, \quad \mathbf{K}_+\vartheta = \chi, \quad \mathbf{K}_-\vartheta = 0. \tag{14}$$

This representation shows similitude to a representation of the Pauli matrices, considering the same restrictions for the $\mathbf{I}_L$ and $\mathbf{I}_R$ minor matrices like above.

### 1.4 *Identifying the group*

All properties of this representation *formally coincide* with a representation of the SU(2) group of the Pauli $\sigma$ matrices. In a similar way, it is a special unitary group with two independent parameters, with the condition that its unitary group element is defined as $\mathbf{E} = \begin{bmatrix} \mathbf{I}_L & 0 \\ 0 & 1 \end{bmatrix}$, (whose $\mathbf{I}_L$ was given in (4) and is a singular minor matrix in the 4 x 4 $\mathbf{E}$). In the latter, restricted sense (considering (6), (8) and (9)) the group formed by the $\tau$-matrices is a special, *pseudo-unitary,* 2 parametric group. We show that this special, pseudo-unitary group is isomorphic with the usual SU(2) group (e.g., among others of the Pauli matrices). The $\tau$-matrices behave like spinor elements of an SU(2) group.

The peculiarities of the $\tau$-matrices (that distinguish them from the Pauli matrices) allow them to make *correspondence between scalars* and *vector components*.

*Theorem 8*: The group of the $\tau$-matrices is *pseudo-unitary* and is *isomorphic* with the special unitary SU(2) group.

*Definition (a)*: An operator $\tau$ in a space $V$ is defined pseudo-unitary if it preserves the form of the scalar product of two vectors: $\langle v\tau, w\tau \rangle = \langle v, w \rangle$, where $\tau \in U(V)$. The pseudo-unitary group $U(n-l, l)$ or $U(V)$ is the group of all operators satisfying this condition. If the scalar product is positive definite (i.e., $l = 0$), then $U(n,0) = U(n)$ is the normal unitary group [35 Chap. 2], [38, Chap. 2.1.5].

*Definition (b):* For $\tau$ is an invertible operator, one can easily check a more simple condition, namely that all $\tau_i$ and $\mathbf{E}$ satisfy that the scalar products be $\langle v\tau, v\tau \rangle = \langle v, v \rangle$ for all $v$ (provided that the vectors $v$ are expressed on an orthogonal 4D basis, when all products $v_iv_j$ vanish, if $i \neq j$). In this case $\tau$ is pseudo-unitary [35, Proposition 2.1, p.68]. In short, according to this condition, the $\tau$ operators *can* form a (special) pseudo-unitary group.

*Proof 1*: In case of the group of our $\tau$ operators, $n = 2$, $l = 0$ or 1. The corresponding groups are either $U(2, 0) = U(2)$ or $U(1,1)$.[11] The latter option may arise in case of a $\tau_3$ representation (whose main diagonal contains two opposite signed elements, i.e., $l = 1$). $U(1,1)$ needs to use an infinite-dimensional Hilbert space, because being a non-compact Lie group $U(1,1)$ does not admit a finite-dimensional unitary representation [30]. Consequently, the pseudo-unitary group of the $\tau$ operators is isomorphic with the unitary U(2) group. As it was shown above, one can assign a |1| absolute value determinant to the (otherwise singular) $\tau$-matrices. In this sense, the $\tau$ operators form a group that is isomorphic with the special unitary SU(2) group. We denote it so in the following.

*Proof 2:* A matrix in *SU*(2) must have the form

---

[11] Ahmed and Jain, [3] showed that a class of group of pseudo-unitary operators is isomorphic to one of the groups $U(n)$ or $U(n,m)$ for some $m,n \in Z^+$.

$$M = \begin{bmatrix} z & w \\ -\bar{w} & \bar{z} \end{bmatrix}$$

where $z$ and $w$ are complex, and $\bar{z}$ denotes the complex conjugate of $z$. To be in $SU(2)$ it is required that $M^+ = M^{-1}$ and that $\text{Det}(M)$ is unitary, where $M^+$ is the conjugate transpose of $M$, and Det denotes determinant. Thus if $z = t_4 + it_3$ and $w = t_2 + it_1$ where $t_4, t_3, t_2, t_1$ are real, and $i^2 = -1$, then

$$M = \begin{bmatrix} t_4 + it_3 & t_2 + it_1 \\ -t_2 + it_1 & t_4 - it_3 \end{bmatrix}$$

with the condition $t_4^2 + t_1^2 + t_2^2 + t_3^2 = 1$.

In our case we can write:

$$M = \begin{bmatrix} (t_4+it_3)\mathbf{I}_L & (t_2+it_1)\mathbf{I}_R \\ (-t_2+it_1)\mathbf{1} & (t_4-it_3)\mathbf{1} \end{bmatrix} = t_1 \begin{bmatrix} & i\mathbf{I}_R \\ i & \end{bmatrix} + t_2 \begin{bmatrix} & \mathbf{I}_R \\ -1 & \end{bmatrix} + t_3 \begin{bmatrix} i\mathbf{I}_L & \\ & -i \end{bmatrix} + t_4 \begin{bmatrix} \mathbf{I}_L & \\ & 1 \end{bmatrix}$$

$$M = t_4 \mathbf{E} + it_1 \tau_1 + it_2 \tau_2 + it_3 \tau_3$$

where $\mathbf{I}_L$ and $\mathbf{I}_R$ are [3x3] chiral unit matrices, $\mathbf{1}$ is [1x1] unit matrix.

In order to meet the condition $t_4^2 + t_1^2 + t_2^2 + t_3^2 = 1$, let's choose $t_1 = t_2 = t_3 = -½$, $t_4 = ½$. Let's introduce the notation $-i\tau_i = \tau_i^q$, where $\tau_i^q$ ($i = 1, 2, 3$) follow quaternion algebra. This means, $(\tau_i^q)^2 = -\mathbf{E}$, $\tau_i^q \tau_j^q = \tau_k^q$ and $\tau_j^q \tau_i^q = -\tau_j^q \tau_i^q$ ($i, j, k$ are cyclic indices), $\tau_i^q \tau_j^q \tau_k^q = -\mathbf{E}$. Then:

$$M = \frac{1}{2}[\mathbf{E} - i\tau_1 - i\tau_2 - i\tau_3] == \frac{1}{2}\left\{\mathbf{E} - \begin{bmatrix} & i\mathbf{I}_R \\ i & \end{bmatrix} - \begin{bmatrix} & \mathbf{I}_R \\ -1 & \end{bmatrix} - \begin{bmatrix} i\mathbf{I}_L & \\ & -i \end{bmatrix}\right\} =$$

$$= \frac{1}{2}\begin{bmatrix} (1+i)\mathbf{I}_L & (1+i)\mathbf{I}_R \\ (-1+i)\mathbf{1} & (1-i)\mathbf{1} \end{bmatrix} = \frac{1}{2}\left\{\mathbf{E} + \begin{bmatrix} & -i\mathbf{I}_R \\ -i & \end{bmatrix} + \begin{bmatrix} & -\mathbf{I}_R \\ 1 & \end{bmatrix} + \begin{bmatrix} -i\mathbf{I}_L & \\ & i \end{bmatrix}\right\}$$

and:

$$M = \frac{1}{2}\left[\mathbf{E} + \tau_1^q + \tau_2^q + \tau_3^q\right]$$

For $\mathbf{E}$, $\tau_i^q$ ($i = 1, 2, 3$) are quaternions, they follow an algebra that excludes complex products. (For the discussion of the comparison of matrix-, quaternion-, and mixed algebras see [15].)

(a) The unitarity of the determinant of $M$ is interpreted in the following form:

$$\text{Det}(M) = \frac{1}{2}\begin{vmatrix} (1+i)\mathbf{I}_L & (1+i)\mathbf{I}_R \\ -(1-i)\mathbf{1} & (1-i)\mathbf{1} \end{vmatrix} = 1 \cdot (\mathbf{I}_L + \mathbf{I}_R)$$

This is a 1 times product of a specially symmetric, unitary formula for the sum of otherwise chiral $\mathbf{I}_L$ and $\mathbf{I}_R$ matrices. $1 \cdot (\mathbf{I}_L + \mathbf{I}_R) = 1 \cdot \begin{bmatrix} 1 & 0 & 1 \\ 1 & 0 & 1 \\ 1 & 0 & 1 \end{bmatrix}$, where the round bracket ( ) contains a 3 x 3 specially symmetric matrix "multiplied" by "1". This unusual unitary form is a condition of the SU(2) compatibility of $M$.

(b) Further, $M^{-1} = \frac{1}{2}\begin{bmatrix}(i-1)\mathbf{I}_L & -(i+1)\mathbf{I}_R \\ (i-1)\mathbf{1} & (i+1)\mathbf{1}\end{bmatrix}$, because

$$M \cdot M^{-1} = \frac{1}{2}\begin{bmatrix}(1+i)\mathbf{I}_L & (1+i)\mathbf{I}_R \\ -(1-i)\mathbf{1} & (1-i)\mathbf{1}\end{bmatrix} \cdot \frac{1}{2}\begin{bmatrix}(i-1)\mathbf{I}_L & -(i+1)\mathbf{I}_R \\ (i-1)\mathbf{1} & (i+1)\mathbf{1}\end{bmatrix} = \begin{bmatrix}1 & 0 & 0 & 0 \\ 1 & 0 & 0 & 0 \\ 1 & 0 & 0 & 0 \\ 0 & 0 & 0 & 1\end{bmatrix} = \mathbf{E}.$$

(c) Adj($M$) = $M^{-1}\cdot$Det($M$). Since $\mathbf{I}_L^2 = \mathbf{I}_L$, $\mathbf{I}_R^2 = \mathbf{I}_R$, $\mathbf{I}_L\mathbf{I}_R = \mathbf{I}_R$ and $\mathbf{I}_R\mathbf{I}_L = \mathbf{I}_L$, there is easy to admit that

$$\text{Adj}(M) = M^{-1}\text{Det}(M) = \frac{1}{2}\begin{bmatrix}(i-1)\mathbf{I}_L & -(i+1)\mathbf{I}_R \\ (i-1)\mathbf{1} & (i+1)\mathbf{1}\end{bmatrix}(\mathbf{I}_L + \mathbf{I}_R)$$

(d) The conjugate transpose of $M$ is $M^+ = \frac{1}{2}\begin{bmatrix}(i-1)(-\mathbf{I}_L) & -(i+1)\mathbf{1} \\ (i-1)(-\mathbf{I}_R) & (i+1)\mathbf{1}\end{bmatrix}$.

In the forms of both $M^{-1}$ (in b) and $M^+$ (in d), the coefficients of ($i$-1) and ($i$+1) are not numbers. Considering that the matrices $\pm \mathbf{I}_L$, $\pm \mathbf{I}_R$, and $\mathbf{1}$ are unitary in their own category, $M^{-1}$ and $M^+$ formally coincide. (Not less than $\tau_i$ and $\sigma_i$ do). □

In summary, $\{\mathbf{E}, \tau_i^q\}$ ($i = 1, 2, 3$) generate a group, whose elements are $\{\pm\mathbf{E}, \pm\tau_i^q\}$. In accordance with the above conditions, these unit quaternions are identified with a representation of an $SU(2)$ compatible quaternion group. Accordingly, $\{\mathbf{E}, \tau_i\}$ ($i = 1, 2, 3$) generate also an $SU(2)$ compatible group, a set of elements of whose representation are $\{\pm\mathbf{E}, \pm\tau_i, \pm i\mathbf{E}, \pm i\tau_i\}$. In both cases the generated groups are pseudo-unitary, and two of their elements are independent.

## 2 Comparing the τ-algebra and the Dirac algebra

### 2.1 *The τ and the Dirac (γ) matrices*

Dirac [20] assigned a set of [4 × 4] $\rho_i$ matrices to the [4 × 4] bispinor $\sigma_i$ matrices by interchanging the second and third rows, and the second and third columns. His $\gamma$ algebra (called generally the Dirac algebra) was defined by the $\gamma_i = \rho_2\sigma_i$ matrices. (Note, there appeared other representations of the Dirac algebra later.)

Notice that the second and third rows and columns of our τ-matrices coincide. Therefore, if we want to assign a set of matrices to the $\tau_i$ matrices – using the analogy of Dirac's $\rho_i$ matrices – they will coincide with the $\tau_i$ matrices themselves. Instead of a $\rho$-$\sigma$ pair, we've got a $\tau$-$\tau$ pair[13].

Consequently, if we define – in a similar way like we obtained the $\gamma$ matrices – a set of T [read: upper case Greek tau] matrices, we get – by definition – the following: $T_i = \tau_2\tau_i$. Since the τ-matrices transform into each other, we get the following algebra:

---

[13] (1) The structure of the expected eigenvectors ($\chi$ and $\vartheta$) determine the structure of the τ-matrices. (2) The analogy to the $\rho$-$\sigma$ matrix set doublet demands that the change of the second rows and columns produce unchanged τ-matrices to achieve identical matrices and avoid duplication, unlike the $\rho$-$\sigma$ pair. (3) $\tau$ is the next letter in the Greek alphabet follwing $\rho$ and $\sigma$ – this justifies the name of the τ-matrices. (4) The rank 2 of the matrices in the τ-algebra is not surprising. These matrices make correspondence between TWO physical quantities. One of these quantities is a *one component* scalar, and the other is a vector – constructed by *three components*. This is reflected in the construction of the τ-matrices.

$$T_1 = \tau_2\tau_1 = -i\tau_3$$
$$T_2 = \tau_2\tau_2 = E \quad \text{(left handed „unit" operator)} \quad (15)$$
$$T_3 = \tau_2\tau_3 = i\tau_1$$

Let us define $T_4 = \tau_4^2 = \tau_3$, where $\tau_4 = \begin{bmatrix} 1 & 0 & 0 & 0 \\ 1 & 0 & 0 & 0 \\ 1 & 0 & 0 & 0 \\ 0 & 0 & 0 & i \end{bmatrix}$

This set defined another representation of the *tau* algebra that we denote by the upper case Greek *T*. The *T*-algebra should behave in a similar way over the *Δ* (isotopic field-charge spin) field, like the *γ* algebra did over the bispinors' space.

Applying the above defined *T*-matrices expressed with *τ*'s, we get the following algebra for the transformation of the eigenvectors $\chi$ and $\vartheta$:

$$\begin{array}{ll} T_1\chi = -i\chi & T_1\vartheta = i\vartheta \\ T_2\chi = \chi & T_2\vartheta = \vartheta \\ T_3\chi = \vartheta & T_3\vartheta = -\chi \\ T_4\chi = \chi & T_4\vartheta = -\vartheta \end{array} \quad (16)$$

Let's introduce the following notations: $\delta_1 = \tau_1^2$; $\delta_2 = \tau_2^2$; $\delta_3 = \tau_3^2$; $\delta_4 = \tau_4^2$.

## 2.2 *The algebra of the δ- and the τ-matrices*

Similar to the K representation of the *τ*-matrices, let's compose an L representation of the *δ*-matrices.

The effects of the $\delta_1 = \delta_2 = \delta_3 = \delta_i = \tau_i^2 = E$ and the $\delta_4 = \tau_3$ matrices on the $\chi$ and $\vartheta$ eigenvectors are:

$$\begin{array}{ll} \delta_i\chi = \chi & \delta_i\vartheta = \vartheta \\ \delta_4\chi = \chi & \delta_4\vartheta = -\vartheta \end{array}$$

Define:

$$L_+ = \frac{1}{2}(\delta_i + \delta_4) = \begin{bmatrix} I_L & 0 \\ 0 & 0 \end{bmatrix} = K_+K_-$$

$$L_- = \frac{1}{2}(\delta_i - \delta_4) = \begin{bmatrix} 0 & 0 \\ 0 & 1 \end{bmatrix} = K_-K_+$$

then:

$$L_+ + L_- = E = \delta_i \text{ and } L_+ - L_- = \tau_3 = \delta_4 \quad \text{(where } i = 1, 2, 3\text{)}$$

The $L_+$ and $L_-$ operators are represented by projector matrices $L_+^2 = L_+$; $L_-^2 = L_-$. Further, in contrast to the **K** operators, they are orthogonal: $L_+L_- = L_-L_+ = 0$. Note, that the $L_+$ and $L_-$ operators are expressed in terms of a $\tau_3$ representation of the tau algebra (by the help of its unit matrix **E** and the $\tau_3$ matrix).

The $\delta_i$ (=**E**) and $\delta_4$ (=$\tau_3$) matrices compose a group, in contrast to the *γ* matrices of the Dirac algebra that do not. The effects of the $L_+$ and $L_-$ matrices on the $\chi$ and $\vartheta$ eigenvectors are:

$$\begin{array}{ll} L_+\chi = \chi & L_+\vartheta = 0 \\ L_-\chi = 0 & L_-\vartheta = \vartheta \end{array}$$

that means, **L**$_+$ acts only on $\chi$ and the orthogonal **L**$_-$ acts only on $\vartheta$. Their effect is the opposite of the effects of the **K**$_\pm$ operators. However, the effects of the $\delta$-matrices and the **L** matrix representation do not modify the ±1/2 eigenvalues of the $\chi$ and $\vartheta$ eigenvectors revealed in Eq. (13).

The effect of the two types of $\delta$-matrices on the eigenvectors $\chi$ and $\vartheta$ of the operators of the $\tau$-algebra can be expressed in the following short [8 x 8] matrix form, where $\delta_i$ and $\delta_4$ are [4x4] matrices that act on the four-element vector columns of $\chi$ and $\vartheta$:

$$\begin{bmatrix} \delta_i & 0 \\ 0 & \delta_4 \end{bmatrix} \begin{bmatrix} \chi & \vartheta \\ \chi & \vartheta \end{bmatrix} = \begin{bmatrix} \chi & \vartheta \\ \chi & -\vartheta \end{bmatrix} \qquad (17)$$

The $\delta_i$ matrices leave intact the eigenvectors $\chi$ and $\vartheta$, while $\delta_4$ changes the sign of one of the eigenvectors, $\vartheta$.

*Application (1) QED:* These $\delta_i$ (i = 1, 2, 3) and $\delta_4$ (together $\delta_\nu$; $\nu$ = 1, ..., 4) matrices extend the Dirac $\gamma$ matrices (cf., Eq. (22)) in an extended Dirac equation in order to restore its covariance lost by the introduction of isotopic field charges in the equation. The $\gamma$ matrices of the original Dirac equation should be replaced by products of the respective $\delta$ and $\gamma$ matrices (cf. Section 3, especially Eqs. (22) and (23)).

*Application (2) Matrix genetics:* S. Petouhov [], [40], [41], [42] referred several times to analogies between the algebra of the genetic code and the algebra applied in QED. The latter matrix transformation confirms those findings. To understand it, please, try to replace the two $\chi$ four-columns in the left of the matrix for the two purine traits (G, A), and the $\vartheta$ four-columns in the right of the matrix for the two pyrimidine traits (C, T) in the DNA bonds in the Eq. (17). In terms of Hadamard matrices [40], [39], this transformation coincides with the transformation of the nitrogenous base Uracil (U) into another nitrogenous base Thymine (T) when one changes from an RNA to DNA during a recombination process, what strictly corresponds to the sign change of the $\vartheta$ in the bottom right position in the 2x2 matrix in the right side of the equation (17). [14]

PART 2

### 3 An example: Application of the $\tau$-algebra in QED, for the extended Dirac equation

In order to demonstrate at least one applicability of the $\tau$-algebra in physics, we present the example of the Dirac equation. The original Dirac equation was invariant under the Lorentz transformation, however, this property did not ensure sufficiency for the covariance of the equation under strongly relativistic conditions. We introduce strongly relativistic conditions in the source of the Dirac equation. This extension is discussed in the following.

---

[14] As (Petoukhov [42], p. 389) writes: An "interesting structural feature of the 8 octets of triplets is connected with the phenomenon of the special status of the T (thymine) in the basic alphabet of DNA. Among the four DNA bases – A, C, G, T – the letter T contrasts phenomenologically with three other letters of the alphabet: 1) only the letter T is transformed into another letter U (uracil) in the transition from DNA to RNA; 2) only the letter T (and its changer U) has not the functionally important aminogroup $NH_2$ in contrast of other three letters … . This binary opposition can be expressed in a digital form". Similar to our separation of the roles of field-charges and velocities in the charge currents, the DNA bases can be asssociated as: A, C and G with a sign + 1 and T with a sign –1. Then "each triplet under replacing its letters on these numbers (A = C = G = + 1, T = –1) can be represented as the product of these numbers. For example, the triplet CAT is represented as 1*1*(–1) = –1 and the triplet TGT - as (–1)*1*(–1) = +1. In the result, the 8 octets of triplets obtain numerical representations as sequences of elements +1 and –1. … The set of these sequences coincide with the complete system of orthogonal Walsh functions for 8-dimensional spaces". The analogy is a sign of general applicability of the $\tau$- and $\delta$-algebras.

The above cited paper [11] showed that $\Delta$ (IFCS) is a two valued, spin-like property that appears in the presence of a kinetic gauge field at high energies. In order to recall [11]: The two isotopic states are associated with the field-charges appearing in the scalar (potential, $V$) and in the vector (kinetic, $T$) parts of a Hamiltonian, respectively.[15] Qualitative distinction was made between the two (isotopic) kinds of field-charges. A given particle can occupy either the one or the other IFC-state.

The wave function of a given particle may be in a potential state with amplitude $\psi_V$, or in a kinetic state with amplitude $\psi_T$.[16,17] The wave function of a single physical particle is an entanglement of the probabilities being in the potential or the kinetic state at a given moment: $\psi = \begin{pmatrix} \psi_T \\ \psi_V \end{pmatrix}$, where $\psi_T$ is a three-component column (cf., footnote 6). We denoted the eigenfunctions $\psi_T$ and $\psi_V$ that belong to the two eigenvalues of the operator $\tau$ by $\varphi_+^{(\tau)}$ and $\varphi_-^{(\tau)}$, respectively. There belong two opposite isotopic field-charge spin ($\Delta$) positions to the sources of the two states. During transition from $\psi_V$ to $\psi_T$ and back, the source of the field, i.e., the respective field-charge, needs to change its $\Delta$ state. The operator $\tau$, whose matrix algebra we have introduced in Section 1 of this paper, affects those $\Delta$ states.

First, recall that $\chi$ and $\vartheta$ are the eigenfunctions of the $\Delta$ isotopic field-charge spin's $\tau$ operators. (As mentioned, the corresponding eigenfunctions were denoted also by $\varphi_+^{(\tau)}$ and $\varphi_-^{(\tau)}$). The eigenvalue equations are $\tau\varphi_+^{(\tau)} = \frac{1}{2}\varphi_+^{(\tau)}$ (or $\tau\chi = \frac{1}{2}\chi$) and $\tau\varphi_-^{(\tau)} = -\frac{1}{2}\varphi_-^{(\tau)}$ (or $\tau\vartheta = -\frac{1}{2}\vartheta$), respectively. The eigenvalues of $\Delta$ can take $\pm\frac{1}{2}$. The operator $\tau$ refers to the transformation matrices of the *group* discussed in Sec. 1.4 that rotate, e.g., the IFCS, in the introduced kinetic gauge field. Since the operators responsible for the rotation in the kinetic gauge field do not affect the space-time dependent and the spin dependent components in the field equations, and vice versa, the particle's full state function $\psi$ can be separated according to the following sum: $\psi = \psi' \varphi_+^{(\tau)} + \psi'' \varphi_-^{(\tau)}$, where $\psi'$ and $\psi''$ denote the state functions affected by the space-time dependent and the spin dependent operators. The $\psi$ state functions in (18), (19) and (21) below are funcions of the $x_v$ space-time coordinates, and of the $r = \pm\frac{1}{2}$ and $s = \pm\frac{1}{2}$ parameters (according to the set of eigenvalues of the $\boldsymbol{\rho}$ and $\boldsymbol{\sigma}$ operators), as well as of the $\Delta = \pm\frac{1}{2}$ parameters (according to the set of eigenvalues of the $\boldsymbol{\tau}$ operator). Let us denote the respective eigenfunctions belonging to the eigenvalues of the $r$, $s$, and $\Delta$ parameters in the state function $\psi$ in the following way: $\varphi_+^{(\rho)}$, $\varphi_-^{(\rho)}$, $\varphi_+^{(\sigma)}$, $\varphi_-^{(\sigma)}$, $\varphi_+^{(\tau)}$, $\varphi_-^{(\tau)}$, according to the $\pm\frac{1}{2}$ eigenvalues of the $\boldsymbol{\rho}$, $\boldsymbol{\sigma}$ and $\boldsymbol{\tau}$ operators. The differential operators act only on $\psi(x_\nu)$, the **σ** operators on the eigenfunctions of $\sigma$, the **ρ** operators on the eigenfunctions of $\rho$, and the **τ** operators on the eigenfunctions of $\Delta$. (To explain this latter notation: please note, we cannot denote the eigenfunctions, which the **τ** operators act on, by $t$, first, because $t$ denotes time, secondly consistence is required with the notations used for the same purpose in [11].)

---

[15] Similar distinction was discussed for the Lagrangians of a system in [27]. They follow a Lagrangian based way to find a Noether symmetry, like I derived in [15], although their derivation of the results seems more complicated, and less general (i.e., holds only for the gravitational interaction).

[16] On partition between the potential and kinetic energies see [45].

[17] According to one of the interpretations of the IFCS theory, the energy of a single particle at a given moment is considered to be ideally concentrated either in the scalar or in the kinetic part of its Hamiltonian [11]. The corresponding state functions belonging to these two idealised reduced Hamiltonians are denoted by $\psi_V$ and $\psi_T$. They characterise the s.c. *potential* and *kinetic* states of the particle. A single particle takes one of these two states at certain *probabilities*. According to this interpretation the particle can change its state (or oscillate) between the fully potential and the fully kinetic states. The observable state function can be characterised by the entanglement of these two probability states and denoted by $\psi$. The IFCS theory describes transition between the two states.

Thus, the $\psi = \psi(x_\nu, r, s, \Delta)$ state function can be dissociated to the following eight products according to the eigenfunctions of the matrix operators:

$$\psi = \psi'\varphi_+^{(\tau)} + \psi''\varphi_-^{(\tau)} =$$
$$\left[\psi_1'(x_\nu)\varphi_+^{(\rho)}\varphi_+^{(\sigma)} + \psi_2'(x_\nu)\varphi_-^{(\rho)}\varphi_+^{(\sigma)} + \psi_3'(x_\nu)\varphi_+^{(\rho)}\varphi_-^{(\sigma)} + \psi_4'(x_\nu)\varphi_-^{(\rho)}\varphi_-^{(\sigma)}\right]\varphi_+^{(\tau)} +$$
$$+ \left[\psi_1''(x_\nu)\varphi_+^{(\rho)}\varphi_+^{(\sigma)} + \psi_2''(x_\nu)\varphi_-^{(\rho)}\varphi_+^{(\sigma)} + \psi_3''(x_\nu)\varphi_+^{(\rho)}\varphi_-^{(\sigma)} + \psi_4''(x_\nu)\varphi_-^{(\rho)}\varphi_-^{(\sigma)}\right]\varphi_-^{(\tau)} = \quad (18)$$
$$\psi_1'(x_\nu)\varphi_+^{(\rho)}\varphi_+^{(\sigma)}\varphi_+^{(\tau)} + \psi_1''(x_\nu)\varphi_+^{(\rho)}\varphi_+^{(\sigma)}\varphi_-^{(\tau)} + \psi_2'(x_\nu)\varphi_-^{(\rho)}\varphi_+^{(\sigma)}\varphi_+^{(\tau)} + \psi_2''(x_\nu)\varphi_-^{(\rho)}\varphi_+^{(\sigma)}\varphi_-^{(\tau)} +$$
$$+ \psi_3'(x_\nu)\varphi_+^{(\rho)}\varphi_-^{(\sigma)}\varphi_+^{(\tau)} + \psi_3''(x_\nu)\varphi_+^{(\rho)}\varphi_-^{(\sigma)}\varphi_-^{(\tau)} + \psi_4'(x_\nu)\varphi_-^{(\rho)}\varphi_-^{(\sigma)}\varphi_+^{(\tau)} + \psi_4''(x_\nu)\varphi_-^{(\rho)}\varphi_-^{(\sigma)}\varphi_-^{(\tau)}$$

This involves also that the solutions of the extended Dirac equation's $\psi$ state function can be separated into eight space-time dependent functions. Opposite to the four solutions of the original Dirac equation, this extended equation has eight solutions, in accordance with the additional bivariant opposite positions of the IFCS. Important to notice: the coincidence of the set of matrices for the $\tau$- and the $T$-algebras guaranties that the number of solutions is *doubled* thanks to the IFCS invariance (according to the two new degrees of freedom brought in by the two possible positions of the IFCS) [18], compared with the *quadruple* solutions of the original Dirac equation due to the non-coincidence of the **ρ** and **σ** operators [20], [21]. It can be exemplified so that, according to the two isotopic states of the field-charges, two separated solutions are assigned to each of the four known solutions of the Dirac equation.

In order to find an algebraic form of the so extended Dirac equation[19], let us start from the source of the extended Dirac equation as derived in [12, Eq. (9)], then discussed in [14], [15]; and reviewed in [13] in the presence of isotopic electric charge densities ($\rho_V$, $\rho_T$) and isotopic gravitational charges ($m_V$, $m_T$) on the one hand, as well as a kinetic gauge field (**D**) on the other hand:

$$\left[-(p_0 + \frac{\rho_T}{c}A_4 + \frac{\rho_V}{c}D_4) - \gamma_5(\boldsymbol{\sigma},\mathbf{p} + \frac{\rho_V}{c}\mathbf{A} + \frac{\rho_V}{c}\mathbf{D}) + \gamma_4 m_V c\right] \cdot$$
$$\cdot \left[(p_0 + \frac{\rho_T}{c}A_4 + \frac{\rho_V}{c}D_4) - \gamma_5(\boldsymbol{\sigma},\mathbf{p} + \frac{\rho_V}{c}\mathbf{A} + \frac{\rho_V}{c}\mathbf{D}) + \gamma_4 m_T c\right]\psi = 0 \quad (19)$$

(The indices $V$ and $T$ refer to the potential and the kinetic members of the isotopic field-charge doublets, namely, in case of the sources of the gravitational field they coincide with the gravitational and inertial masses, respectively, {cf., [11] and [16]}.) Note that $\gamma_4 = \rho_3$, $\gamma_5 = -\rho_1$, and let's apply the *tau* algebra for the eigenvectors of the operators affecting the isotopic field-charges. For simplicity, let us introduce the following expressions for the generalised momenta (in accordance with Dirac's original notations, extended with D):

$$p_4' = -i\left(p_0 + \frac{\rho_T}{c}A_4 + \frac{\rho_V}{c}D_4\right); \quad \mathbf{p}' = \mathbf{p} + \frac{\rho_V}{c}\mathbf{A} + \frac{\rho_V}{c}\mathbf{D}.$$

---
[18] Jentschura bases (cf., [26], [36], [37]) his discussion of the multiple solutions of the Dirac equation to the considerations cited in Footnote 13.

[19] Modifications of the Dirac equation among specific conditions, e.g., [14], present also new solutions for it. Apart from giving quaternionic form to potentials, the two different new solutions provide a remarkable demonstration that non-traditional approches to the electrically charged particle interactions are an actual problem in contemporary physics. Moreover, (Podolsky [44]) discussed its application for half-vectors, showing independence of the choice of *n*-beins, already in 1931.

As shown in [11], the isotopic field charges – in our case both the electric and the gravitational – are rotated by the same transformation matrices in the *isotopic field-charge spin* field[20].[21]

Let's investigate the operators affecting $\psi'$ and $\psi''$ in (18): $\psi = \psi'\varphi_+^{(\tau)} + \psi''\varphi_-^{(\tau)}$ separately. We avoid details of the calculations (one can find them in [15]).

$$\left[i\tau_3\rho_3 p_4' + i\mathbf{E}\rho_2(\boldsymbol{\sigma},\mathbf{p}') + \mathbf{E}m_T c\right]\psi'\chi = \left[i\delta_4\gamma_4 p_4' + i\delta(\boldsymbol{\gamma},\mathbf{p}') + \mathbf{E}m_T c\right]\psi'\chi \quad (20')$$

$$\left[i\tau_3\rho_3 p_4' + i\mathbf{E}\rho_2(\boldsymbol{\sigma},\mathbf{p}') + \mathbf{E}m_T c\right]\psi''\vartheta = \left[i\delta_4\gamma_4 p_4'\chi + i\delta(\boldsymbol{\gamma},\mathbf{p}') + \mathbf{E}m_T c\right]\psi''\vartheta \quad (20'')$$

Thus, after rearrangement, the source of the extended Dirac equation (19) can be written in the following simple form:

$$\left[i\delta_4\gamma_4 p_4' + i\delta(\boldsymbol{\gamma},\mathbf{p}') + \mathbf{E}m_V c\right]\cdot\left[i\delta_4\gamma_4 p_4' + i\delta(\boldsymbol{\gamma},\mathbf{p}') + \mathbf{E}m_T c\right]\psi = 0 \quad (21)$$

We introduce the following notations:

$$\Gamma_1 = \delta_1\gamma_1 = \tau_2 = \begin{bmatrix} 0 & 0 & 0 & -i \\ 0 & 0 & 0 & -i \\ 0 & 0 & 0 & -i \\ i & 0 & 0 & 0 \end{bmatrix}; \quad \Gamma_2 = \delta_2\gamma_2 = -\tau_1 = \begin{bmatrix} 0 & 0 & 0 & -1 \\ 0 & 0 & 0 & -1 \\ 0 & 0 & 0 & -1 \\ -1 & 0 & 0 & 0 \end{bmatrix};$$

$$\Gamma_3 = \delta_3\gamma_3 = \tau_3\rho_2 = \begin{bmatrix} 0 & 0 & -i & 0 \\ 0 & 0 & -i & 0 \\ 0 & 0 & -i & 0 \\ 0 & -i & 0 & 0 \end{bmatrix}; \quad \Gamma_4 = \delta_4\gamma_4 = \mathbf{E} = \begin{bmatrix} 1 & 0 & 0 & 0 \\ 1 & 0 & 0 & 0 \\ 1 & 0 & 0 & 0 \\ 0 & 0 & 0 & 1 \end{bmatrix} \quad (22)$$

Note also that $\Gamma_3 = -i\mathbf{E}\rho_1 = i\mathbf{E}\gamma_5$.

Now, the source of the extended Dirac equation (21) with these notations can be written in the following *more simple* form:

$$\left[i\Gamma_\mu p_\mu' + \mathbf{E}m_V c\right]\cdot\left[i\Gamma_\nu p_\nu' + \mathbf{E}m_T c\right]\psi = 0 \quad (23)$$

The latter is the extended form of the Dirac equation's source in the presence of isotopic field-charges with the application of the tau-algebra. All the rest can be calculated following [11], [12], [14].

---

[20] An approach, similar to our one, was followed by [19]. Deriglazov and Nersessian found – when looking for passages from classical to quantum theory using SO(3.2)-algebra – that quantization of the model led to massive Dirac equation.

[21] The role of the mass term in the Dirac equation, confirming the consistency of quantum mechanics with general relativity, was discussed in detail by [26], that confirmed our assumption to take into account the difference between the gravitaional and inertial masses in the Dirac equation. Jentschura showed the possibility to bring the gravitationally coupled Dirac equation to a form where it can easily be unified with the electromagnetic coupling as it is commonly used in modern particle physics calculations.

Recall again that the transformation of the field-charges in the abstract IFCS field is independent of the given physical interaction (gravitational, electromagnetic, weak, strong); the same transformation rotates the field-charges of all fundamental interactions between their scalar (potential) and vector (kinetic) states appearing in the respective parts of their Hamiltonian. The latter remark holds, of course, also for the field-charges of the electromagnetic and the gravitational fields. The chiral character of the IFCS transformation gets special significance in the case of the electro-weak interacion.

**4 Invariance of the extended Dirac equation**

The (23): $\left[i\Gamma_\mu p'_\mu + \mathbf{E}m_V c\right] \cdot \left[i\Gamma_\nu p'_\nu + \mathbf{E}m_T c\right]\psi = 0$ form of the extended Dirac equation's source demonstrates its covariant character. Two consequences, in comparison with the original Dirac equation, are apparent. Once, the appearance of the opposite isotopic field-charges in the $p_\nu'$ generalised momentum and in the mass terms destroy the Lorentz-invariance of the equation. Secondly, there appears a new invariance, represented by the $\delta$ (or in other notation (22) by the $\tau$) operators in the equation, that rotates the isotopic field-charges in an additional abstract field. The covariance of the extended Dirac equation is resulted in the convolution of these two invariances. The extended Dirac equation is invariant under the combined transformation.

This combined (Lorentz $\otimes$ IFCS) that means (SO$^+$(3,1) $\otimes$ SU(2)) invariance of the equation allows to reinterpret our earlier imagination on the invariances of physical equations. It was a long lasting paradigm in physics that equations of the physical interactions must be subject of invariance under the Lorentz-transformation, and this is not only a necessary, but also a sufficient condition. Now, we see this is not the case. Nevertheless, no physical principle stated that Lorentz-invariance is a sufficient condition demanded for the physical equations. There are other invariances that may appear together and combined with Lorentz's.

PART 3

**5 Additional remarks**

The new invariance described algebraically in this (and introduced in the cited previous) paper(s) is interpreted in the presence of a kinetic gauge field. The Lorentz-invariance is a combined symmetry in itself. The SO$^+$(3,1) group of the proper Lorentz-transformation can be characterised by six independent subgroups. These six independent subgroups can be separated to and characterised by three [4 x 4] rotation matrices [R($\varphi$)] in the space-time, and three [4 x 4] velocity boosts [$\Lambda(\dot{x})$] into a given direction in the configuration space $(= R \otimes \Lambda)$. Since the added IFCS invariance [$\Delta(\dot{x})$] is interpreted also in a kinetic field, it seems reasonable to substitute $R \otimes (\Lambda \otimes \Delta)$ for the $(R \otimes \Lambda) \otimes \Delta$ transformation. Although this clustering is only formal (due to the associativity of group operations), we must mention that the reason is to associate the velocity dependent transformations ($\Lambda$ and $\Delta$) with each other, and formally separate them from the space-time rotation $R$. The latter requires certain intendment change in the approach to the world-picture.

R($\varphi$), $\Lambda(\dot{x}_\nu)$ and $\Delta(\dot{x}_\nu)$ are universal invariance groups, i.e., they concern all fundamental interactions. Nevertheless, the *isotopic field-charge spin*'s SU(2) symmetry [characterised by $\Delta(\dot{x}_\nu)$] is broken at lower energies. At the same time, these invariances are extended by a specific invariance group, characteristic to the given interaction; e.g., in case of electromagnetic interaction by U(1), in case of electroweek interaction by U(1)⊗SU(2), and in case of strong interaction by SU(3).

**Conclusions**

We showed that there exists an invariance group that can transform sources of 3+1 element physical quantities into each other. The group of the $\tau$-matrices can make a unique correspondence between vector components and scalars. This group is isomorphic with the special unitary SU(2) group. This group can make a correspondence between such isotopic physical quantity siblings like the inertial and gravitational masses[22], Lorentz type and Coulomb type electric charges, etc., as predicted analytically in [11], [12], [13], [14] [23] and discussed in [16], and can be applied also in the algebra of the genetic code. The group $\tau$ defines an invariance between particles that compose symmetric pairs. These particle twin siblings differ in their nature and physical properties from those predicted in the SUSY. Therefore, for the sake of distinction, we call their invariance HySy.

**Acknowledgement**

Elaborating first an algebra for the IFC theory was advised by Y. Ne'eman (1925-2006) during the first stage of the project. The project is supported by the Symmetrology Foundation (Budapest), and is developed in the framework of the collaboration „Nonlinear models and symmetry analysis in nonlinear wave phenomena, in the theory of self-organizing systems, as well as in biomechanics, bioinformatics" between the Hungarian and Russian Academies of Sciences (2001-2016).

---

[22] Another approach to distinction between forms of masses is discussed in [8]: According to them a few observed phenomena „suggest that there is a new form of matter that does not shine in the electromagnetic spectrum. Dark matter is not accounted for by either general relativity or the standard model of particle physics. While a large fraction of the high energy community is convinced that dark matter should be described by yet undiscovered new particles, it remains an open question whether this phenomenon requires a modification of the standard model or of general relativity. Here we want to raise a slightly different question namely whether the distinction between modified gravity or new particles is always clear." They showed that this is not always the case.
[23] On the comparision of two descriptions of a Quantum Field Theory, see [49].